\documentclass[12pt]{article}
\usepackage{a4wide}
\usepackage{epsfig}
\newcommand{\oone}{\hbox{$1\kern-2.5pt\hbox{\rm l}$}}
\newcommand{\ssigma}{\hbox{$\kern2.5pt\vrule height4pt\kern-2.5pt\sigma$}}

\newcommand\pfrac[2]{\left(\frac{#1}{#2}\right)}
\newcommand{\Li}{\mathop{\rm Li}\nolimits}
\newcommand{\Cl}{\mathop{\rm Cl}\nolimits}
\newcommand{\imag}{\mathop{\rm Im}\nolimits}

\newcommand{\arcosh}{\mathop{\rm arcosh}\nolimits}
\newcommand{\slasharrow}{\hbox{$\nwarrow\kern-3pt\searrow$}}
\newcommand{\backslasharrow}{\hbox{$\swarrow\kern-3pt\nearrow$}}

\begin{document}
\thispagestyle{empty} 
\begin{flushright}
MITP/14-032
\end{flushright}
\vspace{0.5cm}

\begin{center}
{\Large\bf Virtual and real processes, the K\"all\'en function,\\[.2cm]
 and the relation to dilogarithms}\\[1.3cm]
{\large L.~Kaldam\"ae$^1$ and S.~Groote$^{1,2}$}\\[1cm]
$^1$ Loodus- ja Tehnoloogiateaduskond, F\"u\"usika Instituut,\\[.2cm]
  Tartu \"Ulikool, T\"ahe 4, 51010 Tartu, Estonia\\[7pt]
$^2$PRISMA Cluster of Excellence, Institut f\"ur Physik,
  Johannes-Gutenberg-Universit\"at,\\[.2cm]
  Staudinger Weg 7, 55099 Mainz, Germany
\end{center}

\vspace{1cm}
\begin{abstract}\noindent
We enlighten relations between the K\"all\'en function, allowing in a simple
way to distinguish between virtual and real processes involving massive
particles, and the dilogarithms occurring as results of loop calculations for
such kind of processes.
\end{abstract}

\newpage

\section{Introduction}
The name of the swedish physicist and one of the founders of the CERN, early
deceased Anders Olof Gunnar K\"all\'en (1926--1968), is related forever to
three-particle vertices connecting particles of different masses. As an
example one can use the decay process of the top quark into a bottom quark and
a $W$ boson, with $91\%$ the main decay channel of the top quark seen at the
LHC~\cite{Beringer:1900zz}. Using as a reference frame the rest frame of the
top quark, the kinematics of
\begin{equation}
t(p_t)\to b(p_b)+W^+(p_W)
\end{equation}
can easily be calculated as a warm-up exercise. From four-momentum
conservation one obtains $\vec p_b+\vec p_W=\vec 0$ and $E_b+E_W=m_t$. The
on-shell conditions for the two produced particles read\footnote{If the
particles are not on-shell, the masses for this simple process can be replaced
by off-shell masses.}
\begin{equation}
m_b^2=E_b^2-\vec p_b^2,\qquad m_W^2=E_W^2-\vec p_W^2.
\end{equation}
Inserting $E_W=m_t-E_b$ into the last equation, using
$\vec p_W^2=\vec p_b^2$ and solving for $E_b$ one obtains
\begin{equation}\label{ebw}
E_b=\frac{m_t^2+m_b^2-m_W^2}{2m_t},\quad\mbox{likewise}\quad
E_W=\frac{m_t^2-m_b^2+m_W^2}{2m_t}.
\end{equation}
Finally,
\begin{equation}
\vec p_b^2=E_b^2-m_b^2=\frac{(m_t^2+m_b^2-m_W^2)^2-4m_t^2m_b^2}{4m_t^2}
  =\frac{\lambda(m_t^2,m_b^2,m_W^2)}{4m_t^2}
\end{equation}
where the only mixed product $2m_t^2m_b^2$ changes sign, leading to the
totally symmetric {\em K\"all\'en function}
\begin{equation}
\lambda(a,b,c):=a^2+b^2+c^2-2ab-2ac-2bc.
\end{equation}
Therefore,
\begin{equation}\label{pbw}
|\vec p_b|=|\vec p_W|=\frac{\sqrt{\lambda(m_t^2,m_b^2,m_W^2)}}{2m_t}.
\end{equation}
As will be shown in Sec.~1 of this paper, the K\"all\'en function appears in
different kinematic situations, though describing always the same relation,
namely the realness respectively virtualness of processes involving three
(massive) particles. Parallels to symmetries for dilogarithms are shown in Sec.~2.
Finally, in Sec.~3 we give our conclusions.

\subsection{The K\"all\'en triangle}
\begin{figure}[t]\begin{center}
\epsfig{figure=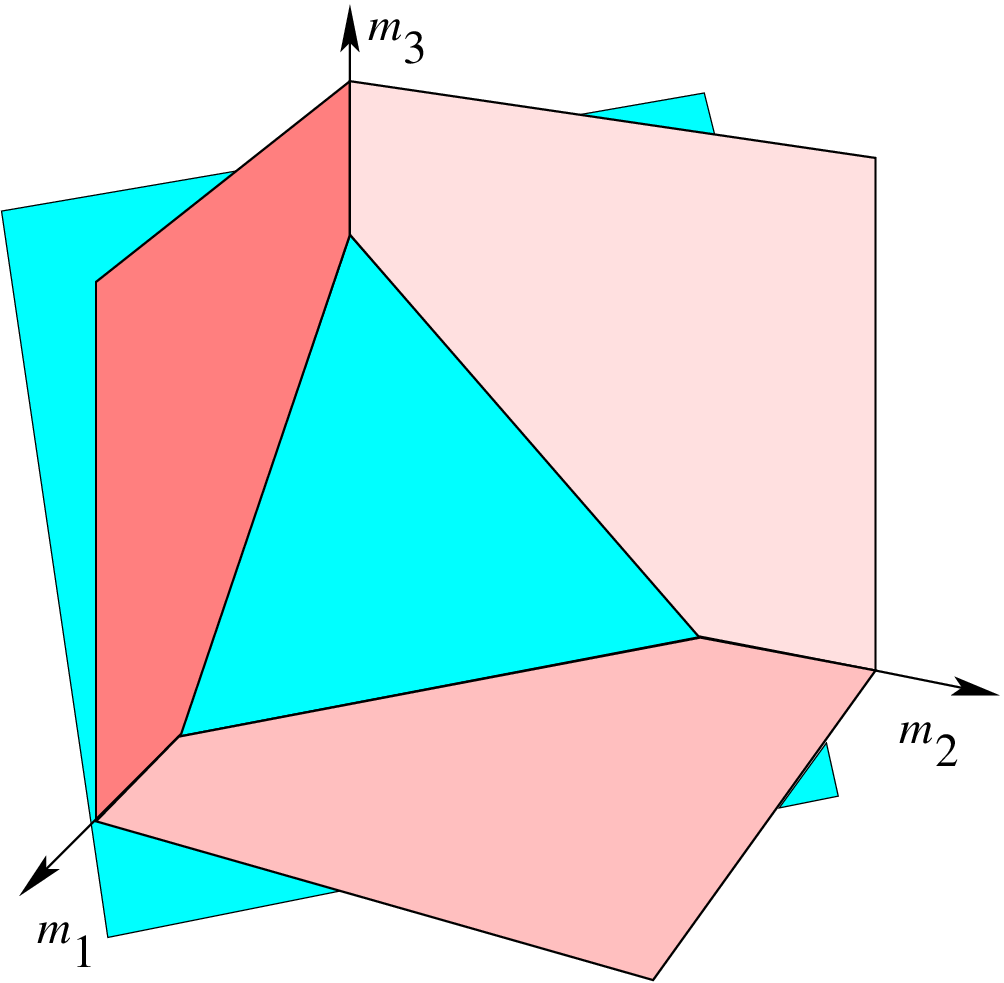, scale=0.5}
\caption{\label{vireoct}Construction of the domain triangle of the
first octant $m_i\ge 0$ ($i=1,2,3$)}
\end{center}\end{figure}

\begin{figure}[t]\begin{center}
\epsfig{figure=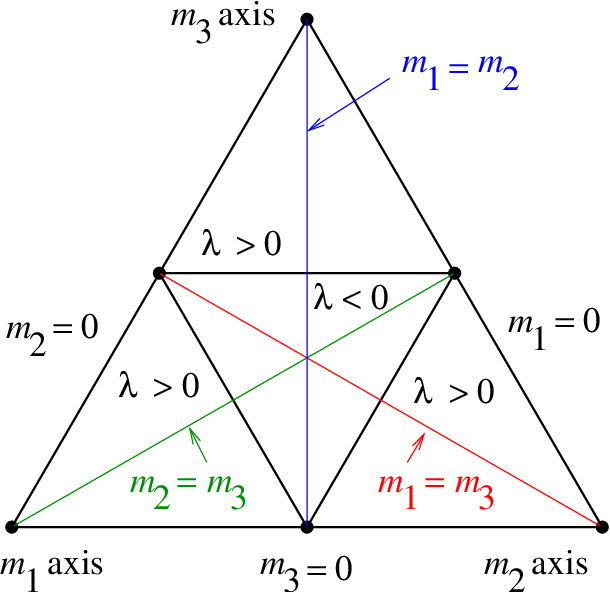, scale=0.8}
\caption{\label{viretri}Regions of positive and negative K\"all\'en function
$\lambda=\lambda(m_1^2,m_2^2,m_3^2)$. The axes of the three-dimensional phase
space are represented by the corners of the domain triangle. The innermost
triangle (with $\lambda<0$) is called the K\"all\'en triangle throughout this
paper.}
\end{center}\end{figure}

Looking more closely at the K\"all\'en function $\lambda(m_1^2,m_2^2,m_3^2)$
of three particles with masses $m_1$, $m_2$ and $m_3$, the analytic behaviour
unfolds a rich spectrum of real and virtual processes. The function is zero
for any one of the thresholds
\begin{equation}
m_1^2=(m_2\pm m_3)^2,\qquad
m_2^2=(m_3\pm m_1)^2,\qquad
m_3^2=(m_1\pm m_2)^2.
\end{equation}
Allowing only positive values for the three masses, in the $(m_1,m_2,m_3)$
phase space the zeros of the K\"all\'en function are located on the three
planes spanned by each two of the three face diagonals $m_1=m_2$, $m_1=m_3$
and $m_2=m_3$. Up to the general normalization one can visualise the phase
space by cutting the first octant by a plane orthogonal to the space diagonal.
On this plane the first octant will appear as upright equilateral {\em domain
triangle\/} (cf.\ Fig.~\ref{vireoct}), the corners representing the $m_i$
axes, and the opposite sides representing the coordinate planes $m_i=0$. The
zeros of the K\"all\'en function are found on the three lines connecting the
midpoints of the sides of the triangle (cf.\ Fig.~\ref{viretri}). The
equilateral triangle enclosed by these three lines is called the
{\em K\"all\'en triangle}. While this triangle is the region where the
K\"all\'en function becomes negative, the other three triangles are regions
with positive K\"all\'en function, each of them containing a coordinate axis.
Special values are
\begin{eqnarray}
\lambda(m_1^2,m_1^2,m_1^2)&=&-3m_1^4\qquad\mbox{(center)},\nonumber\\[7pt]
\lambda(m_1^2,m_2^2,m_2^2)&=&m_1^4-4m_1^2m_2^2\qquad\mbox{(medians)},
  \nonumber\\[7pt]
\lambda(m_1^2,m_2^2,0)&=&(m_1^2-m_2^2)^2\qquad\mbox{(sides)},\nonumber\\[7pt]
\lambda(m_1^2,0,0)&=&m_1^4 \qquad\mbox{(corners)}.
\end{eqnarray}

A negative value of the K\"all\'en function means that the sum of two of the
masses is larger than the third one. This means physically that none of the
three particles can decay into the other two. The process is a virtual one,
leaving at least one of the particles off-shell, i.e.\ with a momentum squared
which is smaller than the squared mass of this particle. On the other hand, a
positive K\"all\'en function means that exactly one decay channel is real.
Taking for instance the region connected to the $m_1$ axis, all points in the
lower left triangle in Fig.~\ref{viretri} represent possible decays through
the channel $1\to 2+3$ because $m_1$ is larger than the sum $m_2+m_3$.
Therefore, one can conclude that the triangle in Fig.~\ref{viretri} gives
structure and direction to processes involving three massive particles.

\subsection{Cascade processes}
Usually, three-particle interactions appear within a more complicated process.
In this case the masses are (partially) replaced by invariant squared momenta
$k_i^2$. On tree level such a process is a cascade process as explained for
instance in Refs.~\cite{Byckling:1970wn,Byckling:1973}. For a cascade process
including $n$ final particles with masses $m_i$ ($i=1,2,\ldots,n$), the
$n$-particle phase space
\begin{equation}
\Phi_n(p;m_1^2,m_2^2,\ldots,m_n^2)=\int\prod_{i=1}^n\frac{d^4p_i}{(2\pi)^4}
  \left(2\pi\delta(p_i^2-m_i^2)\right)(2\pi)^4\delta^{(4)}(p-\sum_{i=1}^np_i)
\end{equation}
can be factorized. Given for instance the momentum $p_1$ of the first
particle emitted from this process, the remaining momentum $k_1=p-p_1$
is mediated by a virtual particle. One can define an invariant mass $M$
with $M^2=k_1^2$ with the only condition that $k_1$ is time-like. Separating
the first phase space integration from the other ones and inserting the
trivial identities
\begin{equation}
1=\int\frac{dM^2}{2\pi}2\pi\delta(k_1^2-M^2),\qquad
1=\int\frac{d^4k_1}{(2\pi)^4}(2\pi)^4\delta^{(4)}(p-p_1-k_1)
\end{equation}
one obtains~\cite{Byckling:1973}
\begin{equation}
\Phi_n(p;m_1^2,m_2^2,\ldots,m_n^2)=\int\frac{dk_1^2}{2\pi}
  \Phi_2(p;m_1^2,k_1^2)\Phi_{n-1}(k_1;m_2^2,\ldots,m_n^2).
\end{equation}
As an example we consider the cascade decay $t\to b+W^+(\to c+\bar s)$.
The first part $\Phi_2(p;m_1^2,k_1^2)$ of the cascade phase space is given by
the phase space
\begin{equation}
\Phi_2(p_t;m_b^2,m_W^2)=\frac{|\vec p_W|}{4m_t(2\pi)^2}\int d\Omega_W
  =\frac{\sqrt{\lambda(m_t^2,m_b^2,m_W^2)}}{8m_t^2(2\pi)^2}\int d\Omega_W
\end{equation}
of the previous (simpler) process $t\to b+W^+$ where $m_W^2$ is replaced by
$k_1^2=p_W^2$. For the second part $\Phi_2(k_1;m_2^2,m_3^2)$ one obtains in a
similar manner
\begin{equation}
\Phi_2(p_W;m_c^2,m_s^2)
  =\frac{\sqrt{\lambda(p_W^2,m_c^2,m_s^2)}}{8p_W^2(2\pi)^2}\int d\Omega_c.
\end{equation}
Assuming that one can perform the solid angle integrations for the $W$ boson
($\Omega_W=4\pi$) and for the $c$ quark ($\Omega_c=4\pi$) trivially, one is
left with the three-particle phase space
\begin{equation}
\Phi_3(p_t;m_b^2,m_c^2,m_s^2)=\int\frac{dp_W^2}{2\pi}
  \frac{\sqrt{\lambda(m_t^2,m_b^2,p_W^2)}}{8m_t^2(2\pi)^2}
  \frac{\sqrt{\lambda(p_W^2,m_c^2,m_s^2)}}{8p_W^2(2\pi)^2}(4\pi)^2.
\end{equation}

\subsection{Limits for the phase space}
The example just introduced is an ideal playground for the implications
caused by the K\"all\'en functions in case of integrations over inner lines.
Even though the intermediate $W$ boson can be off-shell, the phase space has
to be real. Therefore, the two conditions
\begin{eqnarray}
\lambda(m_t^2,m_b^2,p_W^2)&=&((m_t-m_b)^2-p_W^2)((m_t+m_b)^2-p_W^2)\ \ge\ 0,
  \label{kallentbw}\\[7pt]
\lambda(p_W^2,m_c^2,m_s^2)&=&(p_W^2-(m_c-m_s)^2)(p_W^2-(m_c+m_s)^2)\ \ge\ 0
\end{eqnarray}
determine the phase space region for $p_W^2$ and at the same time an
appropriate substitution to calculate the integral analytically. Taking into
account the known quark mass hierarchy, the two conditions result in
\begin{equation}
(m_c+m_s)^2\le p_W^2\le(m_t-m_b)^2.
\end{equation}

\subsection{An appropriate substitution}
Neglecting the masses of $c$ and $s$ quarks, the phase space simplifies to
\begin{equation}\label{intpw2}
\Phi_3(p_t;m_b^2,0,0)=\frac1{512\pi^3m_t^2}\int_0^{(m_t-m_b)^2}
  \sqrt{\lambda(m_t^2,m_b^2,p_W^2)}dp_W^2.
\end{equation}
The square root in Eq.~(\ref{intpw2}) can be simplified by choosing one of the
factors in Eq.~(\ref{kallentbw}) as new variable, e.g.\
$z'=(m_t-m_b)^2-p_W^2$. The second factor is then given by
\begin{equation}
(m_t+m_b)^2-p_W^2=(m_t+m_b)^2-(m_t-m_b)^2+z'=4m_tm_b+z',
\end{equation}
and the upper limit is given by $z'=0$. However, $z'$ is not the optimal
choice. Using instead $z=z'+2m_tm_b=m_t^2+m_b^2-p_W^2$ the square root
simplifies to
\begin{equation}
\sqrt{\lambda(m_t^2,m_b^2,p_W^2)}=\sqrt{z^2-4m_t^2m_b^2}.
\end{equation}
The integration can be performed by using $z=2m_tm_b\cosh\zeta$,
\begin{equation}
\Phi(p_t;m_b^2,0,0)=\frac1{512\pi^3m_t^2}\int_{2m_tm_b}^{m_t^2+m_b^2}
  \sqrt{z^2-4m_t^2m_b^2}dz
  =\frac{m_b^2}{128\pi^2}\int_0^{\zeta_0}\sinh^2\zeta\,d\zeta
\end{equation}
where
\begin{equation}
\zeta_0=\arcosh\pfrac{m_t^2+m_b^2}{2m_tm_b}=\ln\pfrac{m_t}{m_b}.
\end{equation}
Finally, one can use $t=m_b^2e^{2\zeta}$ to obtain
\begin{equation}
\Phi(p_t;m_b^2,0,0)=\frac1{128\pi^2}\int_{m_b^2}^{m_t^2}
  \frac{(t-m_b^2)^2}{8t^2}dt.
\end{equation}

\subsection{Lorentz boosts}
To conclude this section, let us dwell on the parameter $\zeta$ used in the
substitution. This parameter is related to the masses and the invariant
momentum square by
\begin{equation}
\cosh\zeta=\frac{m_t^2+m_b^2-p_W^2}{2m_tm_b},\qquad
\sinh\zeta=\frac{\sqrt{\lambda(m_t^2,m_b^2,p_W^2)}}{2m_tm_b}.
\end{equation}
Comparing with Eqs.~(\ref{ebw}) and~(\ref{pbw}) one realizes that $\zeta$ is
the rapidity for the transition between the $t$ and $b$ quark rest frames, the
exponential representation being
\begin{equation}
e^{\pm\zeta}
  =\frac{m_t^2+m_b^2-p_W^2\pm\sqrt{\lambda(m_t^2,m_b^2,p_W^2)}}{2m_tm_b}.
\end{equation}

\section{Dilogarithms}
Dilogarithms appear in two-fold integrations where the pole of the integrand
is shifted (or in similar arrangements related to this one by the shuffle
algebra~\cite{Broadhurst:1998rz}). Such a shift is usually due to the
occurence of mass contributions. This is the reason why the arguments of the
dilogarithms and their relation to the mass configuration described by the
K\"all\'en function is investigated in this section. The ``classical''
dilogarithm is given by the integral representation
\begin{equation}
\Li_2(z):=-\int_0^z\frac{dz'}{z'}\ln(1-z').
\end{equation}
The classical dilogarithm is a special case of a polylogarithm and is an
analytical function in the whole complex plane except for the interval
$[1,\infty]$ along the positive real axis where the branch cut is located.
Starting from the branch point at $z=1$, the branch cut separates two
different Riemannian sheets of the multi-valued function. There are two
one-parameter and a couple of two- and three-parameter identities that allow
to relate dilogarithms with different arguments~\cite{Lewin:1958,Zagier:1990,%
Bloch:2000,Zagier:2007,Bloch:2013tra}. The two one-parameter identities are
given by~\cite{Lewin:1958}
\begin{eqnarray}
\Li_2(z)+\Li_2\pfrac1z&=&-\frac{\pi^2}6-\frac12\ln^2(-z),\qquad
  z\not\in[0,1[\label{dilog1}\\
\Li_2(z)+\Li_2(1-z)&=&\frac{\pi^2}6-\ln z\ln(1-z).\label{dilog2}
\end{eqnarray}
An additional useful identity is
\begin{equation}
\Li_2(z)+\Li_2(-z)=\frac12\Li_2(z^2).
\end{equation}

\subsection{The hexagon orbit}
For the arguments of the dilogarithms the two identities~(\ref{dilog1})
and~(\ref{dilog2}) are involutions. Because of this, one can create a closed
chain of transitions for the argument which constitutes a hexagon,
\begin{equation}
\begin{tabular}{rcccl}
&&$z$&&\\
&$\nearrow\swarrow$&&$\nwarrow\searrow$&\\
$1-z$&&&&$\displaystyle\frac1z$\\[12pt]
$\uparrow\ \downarrow$&&&&$\uparrow\ \downarrow$\\[7pt]
$\displaystyle\frac1{1-z}$&&&&$\displaystyle1-\frac1z$\\
&$\nwarrow\searrow$&&$\nearrow\swarrow$&\\
&&$\displaystyle\frac{-z}{1-z}$&&\\
\end{tabular}
\end{equation}
For real-valued arguments this transition orbit helps to constrain the
argument to a value lower or equal to $1$, gaining a real value for the
dilogarithm. A pragmatic argument can be very helpful to choose an appropriate
hexagon orbit transition: processes including only real particles have to lead
to real phase space integrals. Therefore, a negative argument for the
logarithms obtained together with the dilogarithm in calculations like
\begin{equation}
\int\ln(\alpha z+\beta)\frac{dz}z=-\Li_2\left(-\frac\alpha\beta z\right)
  +\ln\beta\,\ln z+C
\end{equation}
indicate that the argument of the dilogarithm still is not in an appropriate
form.

\subsection{The Bloch--Wigner dilogarithm}
A similar orbit, though for complex arguments, is given in
Refs.~\cite{Zagier:1990,Bloch:2000,Zagier:2007,Bloch:2013tra} for the
Bloch--Wigner dilogarithm
\begin{equation}
D(z)=\imag\left(\Li_2(z)+\ln|z|\ln(1-z)\right)
\end{equation}
which is mainly (i.e.\ up to a double-logarithmic correction) the imaginary
part of the classical dilogarithm. The identities, induced by the two
involutions, are given by
\begin{equation}
D(z)=-D(z^{-1})=D(1-z^{-1})=-D(-z(1-z)^{-1})=D((1-z)^{-1})=-D(1-z)
\end{equation}
and $D(\bar z)=-D(z)$. In Ref.~\cite{Bloch:2013tra} these identities are
applied to the (complex) root
\begin{equation}\label{zm1m2m3}
z(m_1^2,m_2^2,m_3^2)=\frac{m_1^2-m_2^2+m_3^2
  +\sqrt{\lambda(m_1^2,m_2^2,m_3^2)}}{2m_1^2}
\end{equation}
where the discriminant $\lambda(m_1^2,m_2^2,m_3^2)$ is again the K\"all\'en
function, assumed to be negative.\footnote{For positive values of the
K\"all\'en function the Bloch--Wigner dilogarithm is zero.}
An orbit with constant value for $D$ given by
\begin{equation}
D(z)=D(\bar z^{-1})=D(1-z^{-1})=D(-\bar z(1-\bar z)^{-1})
  =D((1-z)^{-1})=D(1-\bar z)
\end{equation}
is obtained by combining the two involutions with a third one, namely the
calculation of the complex conjugate. As mentioned in
Ref.~\cite{Bloch:2013tra}, the points of the orbit can be obtained from the
starting point $z(m_1^2,m_2^2,m_3^2)$ by interchanging the squared masses
$m_1^2$, $m_2^2$ and $m_3^2$. This invariance describes a symmetry of the
Bloch--Wigner dilogarithm which is manifest already for the K\"all\'en
function. In the following we compare these two functions.

\begin{figure}
\epsfig{figure=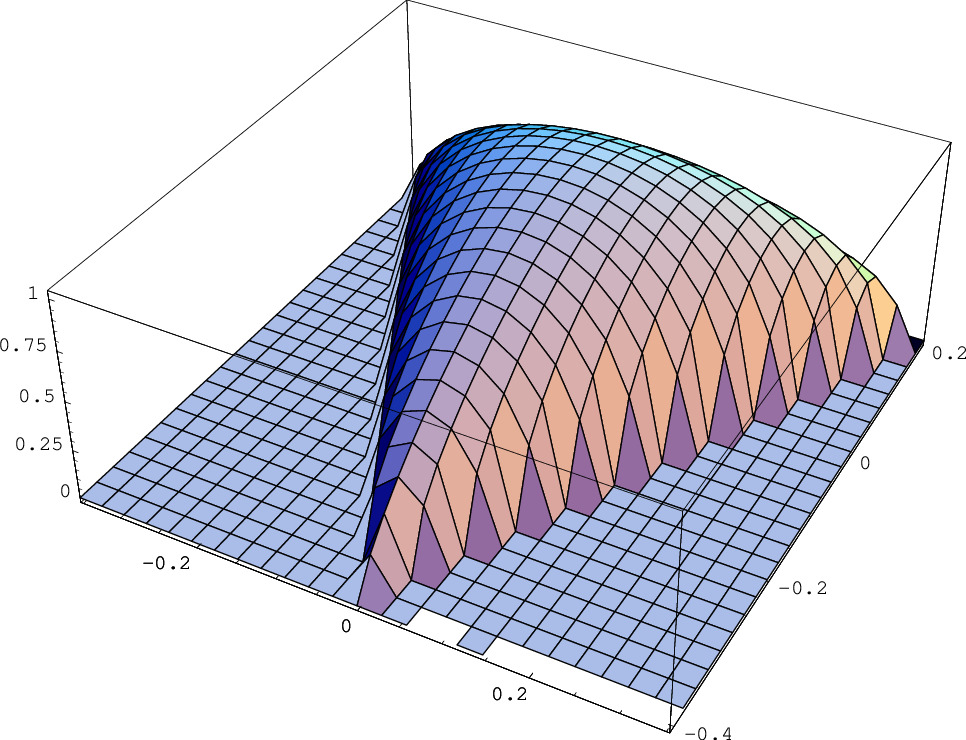, scale=0.8}\\
\epsfig{figure=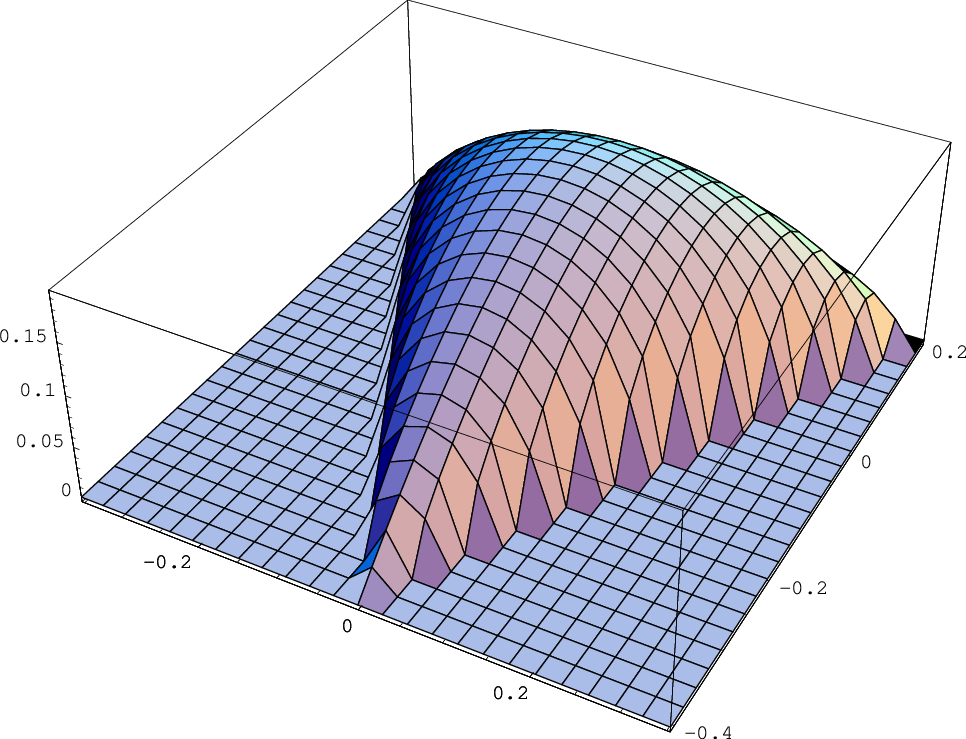, scale=0.8}
\caption{\label{viretop}Topography of $D(z(m_1^2,m_2^2,m_3^2))$,
$z(m_1^2,m_2^2,m_3^2)$ taken from Eq.~(\ref{zm1m2m3}) (top), and
$\imag(\sqrt{\lambda(m_1^2,m_2^2,m_3^2)})$
(bottom) over the K\"all\'en triangle}
\end{figure}

\subsection{Support and similarities}
Because the Bloch--Wigner dilogarithm $D(z)$ vanishes for real values of $z$,
it is evident from Eq.~(\ref{zm1m2m3}) that it vanishes for positive values of
the K\"all\'en function. Therefore, the support of the Bloch--Wigner
dilogarithm and the K\"all\'en function within the domain triangle are both
given by the K\"all\'en triangle shown in Fig.~\ref{viretri}. A first glimpse
at the topographical plot of both functions in Fig.~\ref{viretop} might lead
to the (wrong) conjecture that these functions are the same up to
normalization. Indeed, such a simple relation is impossible because
dilogarithms are transcendental function while the square root of the
K\"all\'en function is not. However, the similarity of these two functions,
if sufficient in the particular application, can be used to give a first
(numerical) estimate of integrals containing the Bloch--Wigner dilogarithm.

\subsection{Parametrizations on the K\"all\'en triangle}
In order to analyse the two functions over the K\"all\'en triangle, instead
of the ``democratic'' parametrization by the three masses $m_1$, $m_2$ and
$m_3$ we use a parametrization which is explicitly two-dimensional. The domain
triangle is parametrized by
\begin{equation}\label{mxmy}
m_x=\frac{m_2-m_1}{\sqrt2(m_1+m_2+m_3)},\qquad
m_y=\frac{2m_3-m_1-m_2}{\sqrt6(m_1+m_2+m_3)}
\end{equation}
and confined by the three straight lines connecting the points
\begin{eqnarray}
(m_x,m_y)&=&\left(\frac1{2\sqrt2},\frac1{2\sqrt6}\right)
  \quad\mbox{for $m_1=0$ and $m_2=m_3$,}\nonumber\\
(m_x,m_y)&=&\left(\frac{-1}{2\sqrt2},\frac1{2\sqrt6}\right)
  \quad\mbox{for $m_2=0$ and $m_3=m_1$ and}\nonumber\\
(m_x,m_y)&=&\left(0,\frac{-1}{\sqrt6}\right)
  \quad\mbox{for $m_3=0$ and $m_1=m_2$.}
\end{eqnarray}
Even though the domain triangle and, as a part of it, the K\"all\'en triangle
in Fig.~\ref{viretri} represent the kinematic situation in the most
symmetrical way (cf.\ Fig.~\ref{viretop}), there is at least a third
representation which is more appropriate for calculating. One can solve the
projective condition $m_1+m_2+m_3=m$ with fixed value $m$ for $m_3$. In this
case the domain triangle is projected onto the $(m_1,m_2)$ plane. The
resulting {\em projective triangle\/} is confined by the two axes and the
line $m_1+m_2=m$.

\subsection{Contour lines}
\begin{figure}[ht]
\epsfig{figure=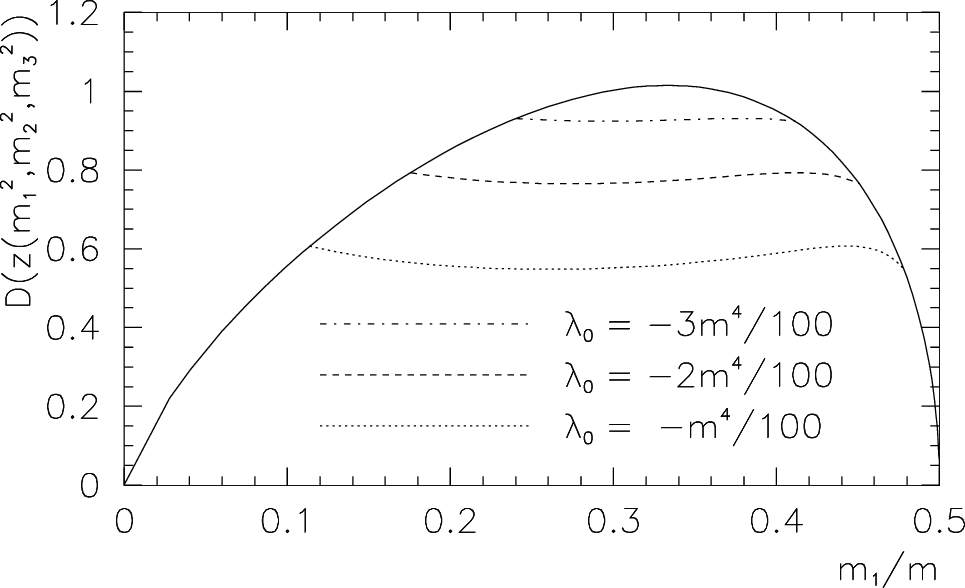, scale=0.95}
\caption{\label{virecon}Bloch--Wigner dilogarithm $D(z(m_1^2,m_2^2,m_3^2))$
(solid line) and curves of constant K\"all\'en value $\lambda_0$ in dependence
on $m_1$, $m_2$ and $m_3$ from Eq.~(\ref{contour})}
\end{figure}

The contour line of the K\"all\'en function with fixed (negative) value
$\lambda_0$ in the projective plane $m_1+m_2+m_3=m$ is given by
\begin{equation}\label{contour}
m_2=\frac12\left(m-m_1\pm\sqrt{m_1^2+\frac{\lambda_0}{m(m-2m_1)}}\right),\qquad
m_3=m-m_1-m_2.
\end{equation}
For $\lambda_0=0$ the contour degenerates to a triangle confined by the lines
$m_1=m/2$, $m_2=m/2$ and $m_2=(m-2m_1)/2$ which is isomorphic to the
K\"all\'en triangle. For $\lambda_0<0$ the contour lines run within this
projective form of the K\"all\'en triangle. It can be easily seen that the
minimal value of the K\"all\'en function is found at $m_1=m_2=m_3=m/3$ with a
value $\lambda_0=-3(m/3)^4=-m^4/27$. In the parametrization~(\ref{mxmy}) of
the domain triangle the minimum is located at $(0,0)$. The value of the
Bloch--Wigner dilogarithm is maximal at the same point $m_1=m_2=m_3=m/3$ (or
$(m_x,m_y)=(0,0)$), resulting in
\begin{equation}
D\left(z\left(\frac{m^2}9,\frac{m^2}9,\frac{m^2}9\right)\right)
  =D\pfrac{1+\sqrt3i}2=D(e^{i\pi/3})=\Cl_2\pfrac\pi3=1.01494\approx 1
\end{equation}
where $\Cl_2(\theta)=\imag(\Li_2(e^{i\theta}))$ is the Clausen function.
However, if taking an intermediate value for $\lambda_0$, a contour line for
the K\"all\'en function has not a constant height for the Bloch--Wigner
dilogarithm. This is shown in Fig.~\ref{virecon} for three different values
of $\lambda_0$.

\section{Conclusions}
In this brief note we have shown that the appearance of the K\"allen function
is always related to the virtuality or reality of three-particle interactions.
We have derived kinematic relations, phase space limits and the relationship
to Lorentz boosts. Starting from the hexagon orbit for classical dilogarithms
we have considered the invariance of the Bloch--Wigner dilogarithm under a
similar hexagon orbit, showing similar symmetry properties and the same
support for the Bloch--Wigner dilogarithm and the imaginary part of the
square root of the K\"all\'en function. In addition, we have found a rough
approximation for the (transcendental) Bloch--Wigner dilogarithm given by
\begin{equation}
D(z(m_1,m_2,m_3))\approx
  \sqrt{27}\imag\left(\sqrt{\lambda(m_1^2,m_2^2,m_3^2)}\right)
\end{equation}
which can be used for a first estimate of integrals including the
Bloch--Wigner dilogarithm.

\subsection*{Acknowledgements}
This work was supported by the Estonian Institutional Research Support under
grant No.~IUT2-27, and by the Estonian Science Foundation under grant No.~8769.
S.G.\ acknowledges support by the Mainz Institute of Theoretical Physics
(MITP).

\end{document}